[2023/10/19]
\documentclass[draft]{article}
\title{Probability, Curvature and Spectrum on Graphs}
\author{Tianhong Zhao\\{\footnotesize University of Science and Technology of China}\\{\footnotesize Email: \texttt{ustczth@mail.ustc.edu.cn}}}
\date{2026/7/14}

\usepackage{amsfonts}
\usepackage{amsmath}
\usepackage{amssymb}
\usepackage{mathtools}
\usepackage{amsthm}
\usepackage{pgfplots}
\usepackage{enumitem}
\usepackage{tikz} 
\usepackage{tikz-cd}
\usepackage{changepage}

\chardef\bslash=`\\ % p. 424, TeXbook
\newtheorem{thm}{Theorem}[section]

\newtheorem{prop}[thm]{Proposition}

\theoremstyle{definition}
\newtheorem{defn}{Definition}[section]

\theoremstyle{remark}
\newtheorem{rem}{Remark}[section]

\newcommand{\R}{\mathbb{R}}
\newcommand{\Z}{\mathbb{Z}}

\begin{document}
	\maketitle
\begin{abstract}
We show that, for a metric graph equipped with the Laplacian operator $\Delta=-\frac{d^2}{dx^2}$, the graph trace formula admits a new interpretation in terms of quantum probability and curvature. Our approach is based on a notion of graph curvature inspired by Wilson-lines and holonomy, together with von-Neumann's ergodic theorem.
\end{abstract}

\section{Introduction}
It is well-known that graph trace formula associate the primitive path on graph with the spectrum of graph. Let $G$ be a undirect finite graph, the Ihara zeta function is: 
$$
\zeta_{\text{Ihara}}(u)=\frac{1}{\det\left(1-Au\right) }=\prod_{\gamma}\frac{1}{1-u^{\ell(\gamma)}}=\prod_{\lambda}\frac{1}{1-\lambda u},
$$
where $u$ is a indeterminate, $A$ is the edge adjacency matrix of $G$, $\gamma$ is the primitive closed path on graph, $\ell(\gamma)$ is the length of $\gamma$, $\lambda$ is the graph spectrum of $A$.

Such construction could be promoted to a finite metric graph with a Laplacian operator $\Delta=-\frac{d^2}{dx^2}$ (also called quantum graph), suppose the Kirchhoff boundary conditions hold, suppose every edge has a length $1$. It is known that:
\begin{thm}(Trace formula on quantum graph\textup{\cite{Kurasov2024}})
$$
\frac{1}{2\pi }\sum_{n\in \Z} Tr\left(U^{n}\left(k\right) \right) =\frac{1}{2\pi}\sum_{\gamma}\ell \left(prim\left( \gamma\right)  \right) S_{\gamma}e^{ik\ell\left(\gamma\right)}=\sum_{k_0} \delta\left(k-k_0\right), 
$$
\end{thm}

where $k$ is a real parameter represents energy, $U\left(k\right)$ is the graph scattering matrix of a quantum graph, $\gamma$ is the primitive closed path on graph, $\ell\left(\gamma \right)$ is the length of path,  $S_{\gamma}$ is the scattering coefficient of $\gamma$, $k_0$ is the spectrum of $\Delta$.

An early trace formula for the Laplacian on metric graphs was derived by Roth \cite{Roth1984} in 1984. A scattering-theoretic approach was subsequently developed by Kottos and Smilansky \cite{KottosSmilansky1999} in 1999, who established an exact trace formula relating the quantum spectrum to periodic orbits on the graph. In 2005, Kurasov and Nowaczyk \cite{KurasovNowaczyk2005} gave a rigorous derivation of the trace formula, refining its earlier formulations. Bolte and Endres \cite{BolteEndres2009} subsequently extended the formula to more general self-adjoint vertex conditions. Rueckriemen–Smilansky developed a graph trace formula with edge potentials. In 2024, Kurasov \cite{Kurasov2024} published a book  about trace formula on quantum graph. We refer to \cite{Kurasov2024} as our main reference.

Our new perspective is the trace formula on quantum graph could be modified as:
$$
\textbf{Quantum Probability}=\textbf{Graph Curvature}=\textbf{Energy Spectrum}
$$
The key observations of trace formula on quantum graph are:
\begin{enumerate}
	\item  The left term could be explained as a Quantum Probability. From Von-Neumann's ergodic theorem, we know:	$$\lim_{N\to \infty}\frac{1}{2N}\sum_{m=-N}^{N}U^{m}=P,$$
	where $P$ is the projection on invariant subspace.
	
	So if we consider a density matrix $\rho$, 
	$$Tr( P\rho)=\lim_{N\to \infty}\frac{1}{2N}\sum_{m=-N}^{N} Tr\left(U^{n}\left(k\right)\rho\right)$$
	is the probability of $P$.
    \item The middle term consists of closed primitive paths. $Tr\left(U^{n}\left(k\right)\rho\right)$ could be explained as Graph Curvature defined by Wilson-line and holonomy. It is a parallel transport of $\rho$ in the average sense.
    \item The right term is the Energy spectrum of $\left(G,\Delta\right)$.
\end{enumerate}

Let $G$ be a finite metric graph with discrete length, $\Delta=-\frac{d^2}{dx^2}$ be the graph Laplacian. The main theorem of this paper is:
\begin{thm}$\left( \textbf{Probability}=\textbf{Curvature}=\textbf{Spectrum}\right) $
$$
	\begin{aligned}
		\frac{1}{2\pi}
		\sum_{n\in\mathbb{Z}}
		Tr\left(U^{n}(k)\rho(k)\right)
		&=
		\frac{1}{2\pi}
		\sum_{\gamma}
		S_{\gamma,\rho}e^{ik\ell(\gamma)}
		\\
		&=
		\sum_{k_0}
		\left\langle
		\psi(k_0)
		\middle|
		\rho(k_0)
		\middle|
		\psi(k_0)
		\right\rangle
		\delta(k-k_0),
	\end{aligned}
$$
\end{thm}
where $k$ is a real parameter represents energy, $U\left(k\right)$ is the graph scattering matrix of $\left(G,\Delta\right)$, $\rho$ is density matrix, $\gamma$ is oriented path on graph, $\ell\left(\gamma \right)$ is the length of path, $S_{\gamma,\rho}$ is the scattering coefficient of $\gamma$, $k_0$ is the spectrum of $\Delta$, $\left|\psi\left(k_0\right) \right\rangle$ is the eigenvector of $U\left(k_0\right)$.

Our graph trace formula has the potential to establish a new connection between quantum information and geometry. The underlying construction is conceptually reminiscent of the Quantum probability distribution and the geometric notions of Wilson lines and holonomy.

\section{Quantum Probability}
Let $\rho$ be a density matrix on a Hilbert space $\mathcal{H}$, $P$ be an orthogonal projection. Thus we have: $P^{\dagger}=P,P^2=P$. The quantum measurement with respect to $P$ is defined as:
$$
\rho\mapsto \rho^{*}=\frac{P\rho P}{Tr\left(P\rho P\right)},
$$
where $\rho^{*}$ is the density matrix after quantum measurement, 
$$
\Pi= Tr\left(P\rho P\right)=Tr\left( P^2\rho\right)=Tr\left(P\rho \right)
$$
is the probability of $P$. Suppose we have an orthogonal projection family $\left\lbrace P_1,P_2,...\right\rbrace$ satisfy:
$$
P_iP_j=0 \ \ \forall i\ne j;\  \sum_{k} P_k=\textbf{I},
$$
where $\textbf{I}$ is the identity of Hilbert space $\mathcal{H}$.

Then the normalization condition holds:
$$
\Pi_k= Tr\left(P_k \rho \right), \ \sum_{k} \Pi_k=Tr\left(\rho \sum_{k}P_k\right)=Tr\left(\rho\right)=1.
$$
\begin{defn}
	We call $\left\lbrace \Pi_k\right\rbrace_{k=1,2,...}$ \textbf{Quantum Probability} respect to the density matrix $\rho$ and orthogonal projection family $\left\lbrace P_k\right\rbrace_{k=1,2,...}$.
\end{defn}
The Quantum Probability $\Pi_k$ is real and positive.
\section{Graph Curvature}
It is known that graph curvature could be realized as the Ollivier--Ricci curvature by using the theory of optimal transport, which is a very successful theory. We discuss this definition in the section of future directions. There are some other kinds of graph curvature. We use a definition of graph curvature which is similar as Wilson-line and holonomy different from Ollivier--Ricci curvature. In differential geometry, curvature could be defined as the change of vector after a parallel transport along with a loop $\gamma$. This definition could be written as a linear transform on the tangent space:
$$
P_{\gamma}:T_pM\to T_pM,\ \  P_{\gamma}=\mathcal{P}\exp\!\left(-\int_{\gamma}\omega\right),
$$
where $\mathcal{P}$ denotes path ordering, $\omega$ is the connection 1-form on manifold $M$, $P_{\gamma}$ is the parallel transport operator along the path $\gamma$.

Let $(G=(V,E))$ be an undirected graph, $A_G$ be the vertex adjacency matrix of $G$. When it comes to a graph, it is convenient to consider all the cycles of $G$ and define the parallel transport alone the cycle. We have:
$$
Tr(A_G^n)=\text{number of cycles of length $n$}.
$$
$A_G$ as a parallel transport operation. Noted that the Ricci curvature $R_{ij}$ is the average of Riemann curvature, we have:
$$
R_{ij} \sim \sum_{n=1}^{\infty}Tr(A_G^nE_{ij})\sim \text{number of paths from vertex $i$ to vertex $j$},
$$
where $E_{ij}$ is a standard basis matrix which has $1$ at the $\left(i,j\right)$-th entry and $0$ otherwise. Let $\rho$ be a density matrix. We view the density matrix as a vector, it is possible to consider the average parallel transport of $\rho$ \footnote{It could be also viewed as graph path partition function.}:
$$
P(\rho)=\sum_{n=1}^{\infty}Tr(A_G^n\rho)
$$

This roughly definition is related to Wilson-line and holonomy but it has several drawbacks:
\begin{enumerate}
	\item This definition relied on the globally spectrum of graph, not a locally definition.
	\item Paths on graph should have orientation. The vertex adjacency matrix $A_G$ could give a wrong contribution to Ricci curvature.
\end{enumerate}

We would study $1$ as our future direction.

We use the \textbf{Graph Scattering Matrix} to solve $2$, which is a quantization of edge adjacency matrix. 

Let $G$ be a finite metric graph with laplacian operator $\Delta=-\frac{d^2}{dx^2}$. We view the graph as a gluing of intervals, say $\left[0,1\right]_j,j=1,2,...N$. Every interval  correspond to an edge $E_j$. We study the eigenvalue problem of $\Delta$ together with Kirchhoff boundary conditions.

For an edge $E_j$, we have:
$$
\Delta f =-\frac{d^2}{dx^2}f =k^2f
$$
The solution on each edge $E_j$ is:
$$f(x)=A_{2j-1,in}e^{ikx}+B_{2j-1,out}e^{-ikx},$$
where $A_{2j-1,in}$ is the in-coming wave and $B_{2j-1,out}$ is the out-going wave. If we use the other vertex as an origin, the solution is: 
$$f(x)=A_{2j,in}e^{-ik\left(x-1\right) }+B_{2j,out}e^{ik\left(x-1\right) }.$$
We could define a transfer matrix $S_{e}\left(k\right)=\begin{pmatrix}
	0& e^{ik} \\
	e^{ik} &0
\end{pmatrix}$:
$$
\begin{pmatrix}
	A_{2j-1,in} \\
	A_{2j,in}
\end{pmatrix}=\begin{pmatrix}
	0& e^{ik} \\
	e^{ik} &0
\end{pmatrix}\begin{pmatrix}
	B_{2j-1,out} \\
	B_{2j,out}
\end{pmatrix},
$$
Let's define:
$$
\vec{A}_{in}=\begin{pmatrix}
	A_{1,in} \\
	A_{2,in} \\
	... \\
	A_{2N,in}
\end{pmatrix},
\vec{B}_{out}=\begin{pmatrix}
B_{1,out} \\
B_{2,out} \\
... \\
B_{2N,out}
\end{pmatrix}.
$$
Then we have:
$$
\vec{A}_{in}=\begin{pmatrix}
	0 & e^{ik} & 0 &  0&... \\
	e^{ik} & 0 & 0 &  0& ...\\
	0 & 0 & 0 & e^{ik}  &... \\
	0 & 0 & e^{ik} & 0  &... \\
	... & ... &...  &...  &...
\end{pmatrix}\vec{B}_{out}=S_e\left(k\right)\vec{B}_{out},
$$
where the transfer matrix:
$$
S_e\left(k\right)=\begin{pmatrix}
	0 & e^{ik} & 0 &  0&... \\
	e^{ik} & 0 & 0 &  0& ...\\
	0 & 0 & 0 & e^{ik}  &... \\
	0 & 0 & e^{ik} & 0  &... \\
	... & ... &...  &...  &...
\end{pmatrix}
$$ 
is called \textbf{Edge Scattering Matrix}.

For a vertex $p$, Kirchhoff boundary conditions are defined as:
$$
\begin{cases}
	f(c_1)=f(c_2)=...=f(c_d)\\
	\sum_{i=1,...d} {f}' (c_d)=0,
\end{cases}
$$
where $c_1,c_2...c_d$ represents vertex of interval connected to $p$, the derivatives are taken in the directions away from the vertex $p$. It can be shown that these conditions\cite{Kurasov2024} could be written as:
$$
S_{v,p}=\begin{pmatrix}
	\frac{2-d}{d} & \frac{2}{d} & \frac{2}{d} & \frac{2}{d} & \cdots \\
	\frac{2}{d} & \frac{2-d}{d} & \frac{2}{d} & \frac{2}{d} & \cdots \\
	\frac{2}{d} & \frac{2}{d} & \frac{2-d}{d} & \frac{2}{d} & \cdots \\
	\frac{2}{d} & \frac{2}{d} & \frac{2}{d} & \frac{2-d}{d} & \cdots \\
	\vdots & \vdots & \vdots & \vdots & \ddots
\end{pmatrix},
$$
where $S_{v,p}$ represents vertex boundary conditions on the vertex $p$.
\begin{defn}
We associate each undirected edge with a pair of oppositely oriented edges. Let $E,E'$ be directed edges of $G$, $\bar{E}$ be the opposite directed edge of $E$, $d_p$ be the valence of $p$. The $2N\times 2N$ matrix: $$
\left( S_v\right)_{E',E} =\begin{cases}
	\frac{2}{d_p}-\delta_{E',\bar{E}} &  E,E'\ \text{join at}\ p,\\
	0 & \text{ otherwise}   
\end{cases}
$$
is called the \textbf{Vertex Scattering Matrix}, where $p$ runs over all vertex of $G$, $\delta_{E',\bar{E}}$ is Kronecker delta.
\end{defn}
$S_v$ relates the in-coming and out-going waves at the vertex and is unitary under the Kirchhoff boundary conditions. We have:
$$
\vec{B}_{out}=S_v\vec{A}_{in},
$$
hence the following holds:
$$
\vec{A}_{in}=S_e(k)\vec{B}_{out}=S_e(k)S_v\vec{A}_{in}.
$$
Let $U\left(k\right)=S_e(k)S_v$, then $k \in \R$ is eigenvalue of Laplacian $\Delta$ if and only of $1$ is eigenvalue of $U\left(k\right)$, the eigenvalue of $\Delta$ is given by the secular equation: 
$$\det\left(U\left(k\right)-I\right)=0.$$
The unitary matrix family $U\left(k\right)$ is called \textbf{Graph Scattering Matrix} $U\left(k\right);k\in \R$.

The graph scattering matrix $U\left(k\right)$, which contains all the distance and topology information of a metric graph, is a quantization of edge adjacency matrix.

From Von-Neumann's ergodic theorem, we know:
$$
\lim_{N\to \infty}\frac{1}{2N}\sum_{m=-N}^{N}U^{m}\left(k\right) =P\left(k\right),
$$
$P\left(k\right)$ is the projection of $U\left(k\right)$ with eigenvalue $1$, $P\left(k\right)=0$ except for the spectrum of $\Delta$. Let $\rho\left(k\right) $ be the density matrix, take the trace we have:
\begin{prop}\textbf{(Quantum Probability=Graph Curvature)}
$$
\Pi_k=Tr\left( P\left(k\right)\rho\left(k\right)\right)=
\lim_{N\to \infty}\frac{1}{2N}\sum_{m=-N}^{N}Tr\left( U^{m}\left(k\right)\rho\left(k\right)\right)
$$
\end{prop}

The left hand side is the \textbf{Quantum Probability} while the right hand side is our definition of \textbf{Graph Curvature} (the parallel transport is given by $U\left(k\right)$). \\

\begin{rem}
$U\left(k\right)$ is defined on the linear space $\mathcal{H}$ spanned by the coefficient $\left\lbrace A_{j,in}\right\rbrace_{j=1,2,...2N}$. 

It might be misunderstood that whether $P\left(k\right)$ could represent the spectral measure of graph Laplacian $\Delta$ and what is the normalization condition. We shall clear this statement.
Recall that the eigenvalue function $f_k \in L^2(G)$ is given by the equation:
$$
\Delta f_k=k^2 f_k.
$$
Let's define a linear transform $\mathcal{T}_k:\mathcal{H}\to L^2(G)$,
$$\left(\vec{A}_{in}\right)\mapsto f\left(x\right)\mid_{x\in E_j} =A_{2j-1,in}e^{ikx}+B_{2j-1,out}e^{-ikx},$$ where $f\left(x\right)$ is piecewise function on each edge $E_j$. For every $k\in \R$, 
the projection $P\left(k\right)$ on $\mathcal{H}$ should be coincide as the spectral measure $\widetilde{P\left(k\right)}$ on $L^2(G)$ via linear transform $\mathcal{T}_k$. It follows that this diagram of Hilbert space commutes:
$$
\\
\begin{tikzcd}
	\mathcal{H} \arrow["P\left(k\right)"',d]  \arrow["\mathcal{T}_k"',r] & L^2(G) \arrow["\widetilde{P\left(k\right)}"',d]\\
	\mathcal{H} \arrow["\mathcal{T}_k"',r]&  L^2(G)
\end{tikzcd}
\\
$$
It is easy to check $\mathcal{T}_k$ is injection for $k\ne 0$, so the projection on $\mathcal{H}$ could be transferred faithfully to spectral measure on $L^2(G)$. If $k=0$, $\mathcal{T}_0$ may have non-trivial kernel and gives an extra topological information of $G$. It would gives a correction of $0$-energy term in graph trace formula but it is not directly related to our results. We refer to Kurasov's book for the precisely proof.

Due to the completeness of graph Laplacian spectrum, we have $\left\lbrace \widetilde{P\left(k\right)},\ k\in \R \right\rbrace $ is normalized and orthogonal. It is just the spectral decomposition of $\Delta$ act on $L^2(G)$. The density matrix $\rho\left(k\right) $ should be coincide with the density matrix $\widetilde{\rho}$ on $L^2(G)$:
$$
\Pi_k=Tr\left( P\left(k\right)\rho\left(k\right)\right)=Tr\left( \widetilde{P\left(k\right)}\widetilde{\rho}\right), \sum \Pi_k=1.
$$
\end{rem}

The connection between \textbf{Graph Curvature} and \textbf{Energy Spectrum} of $G$ is given by trace formula on quantum graph:
$$
\frac{1}{2\pi}\sum_{m\in \Z}Tr\left(U^m\left(k\right) \right)= \sum_{k_0} \delta\left(k-k_0\right), 
$$
where $k_0\in \R$ satisfy: $\det\left(U\left(k_0\right)-I\right)=0$, the set of $k_0$ is called \textbf{Energy Spectrum} of quantum graph $G$.  $\delta\left(k-k_0\right)$ are Dirac delta functions\footnote{The equation $\det\left(U\left(k\right)-I\right)=0$ may have non-trivial algebraic multiplicity when $k_0=0$, it could give a correction of $\delta\left(k\right)$. The coefficient of  $\delta\left(k\right)$ is equal to $Tr\left(P(0)\right)$.}.

This formula could be understood as an identity of a unitary matrix family $U\left(k\right)$. If we calculate some example of quantum graph with discrete length, this would become a finite combine of Poisson's summation formula. If the lengths of edge are rational independent, The trace formula would transfer into a crystalline measure \cite{Kurasov2024}\cite{cry}.

In quantum graph theory, $Tr\left(U^{m}\left(k\right)\right)$ could be calculated by the closed primitive path which has a discrete length $m$. It can be shown that:
$$
Tr\left(U^{m}\left(k\right)\right)=\sum_{\gamma ,\ell\left(\gamma\right)=m}\ell \left(prim\left( \gamma\right)  \right) S_{\gamma}e^{ik\ell\left( \gamma\right)}, 
$$
where sum runs over all closed primitive paths on with discrete length $m$, $\ell \left(prim\left(\gamma\right)  \right)$ is the length of primitive path, $S_{\gamma}$ is the scattering coefficient of the path.

Noted that in usual trace formula, we only know the close primitive paths on graph. Due to the density matrix $\rho\left(k \right) $, we know the information of open paths on graph. It can be shown that:
$$
Tr\left(U^{m}\left(k\right)\rho\left(k\right)\right)=\sum_{\gamma ,\ell\left(\gamma\right)=m} S_{\gamma,\rho}e^{ik\ell\left( \gamma\right)}, 
$$
where sum runs over all oriented paths with discrete length $m$, $\rho\left(k\right)$ is the weight of summation, $S_{\gamma,\rho}$ is the scattering coefficient of the path.
The cost is that we should have the information of eigenvector of $U\left(k\right)$ or probability of quantum state. My previous preprint \cite{Zhao2025} on arXiv showed that the following identity holds:
\begin{thm}\textbf{(Graph Curvature=Energy Spectrum)}
$$
\frac{1}{2\pi}\sum_{m\in \Z}Tr\left(U^m\left(k\right)\rho\left(k \right) \right)= \sum_{k_0} \left \langle \psi\left (k_0\right )  |\rho\left(k_0 \right) |\psi\left (k_0\right )   \right \rangle \delta\left(k-k_0\right),
$$
where $k_0\in \R$ satisfy: $\det\left(U\left(k_0\right)-I\right)=0$, $\left| \psi\left(k_0\right) \right\rangle$ is the eigenvector of $U\left(k_0\right)$.
\end{thm}

As a conclusion, we get the following correspondence:
$$
\textbf{Quantum Probability}=\textbf{Graph Curvature}=\textbf{Energy Spectrum}
$$

\section{Future direction}
Recall the Einstein equation is: 
$$
\textbf{Ricci}-\frac{1}{2}\textbf{Rg}=\textbf{T},
$$
In a slogan:
$$
\textbf{Curvature}=\textbf{Energy-Momentum tensor}
$$

It remains unclear how to generalize this result to Lorentzian spacetime and how to recover local curvature quantities, such as the Ricci and scalar curvatures, from the global curvature encoded by Wilson-line and holonomy. Recently, \cite{barton2026ollivierriccicurvaturecausalsets}\cite{MondinoSuhr2023}  new results in Lorentzian geometry shows timelike Ricci curvature can be recovered up to higher-order terms from optimal transport of probability measure. It may provide a promising direction\footnote{These probability measure are defined over real field.}:
$$
	\ell_1(\mu_x,\mu_y)
	=
	\delta
	\left[
	1+
	\frac{\varepsilon^2}{2}
	\frac{n}{(n+1)(n+2)}
	\operatorname{Ric}(v,v)
	+
	O\!\left(\varepsilon^3+\varepsilon^2\delta\right)
	\right],
$$
where $\ell_1(\mu_x,\mu_y)$ is $\ell_1$-Lorentzian--Wasserstein distance, $\varepsilon$ is the diagram of causal diamond, $y=\exp_{x}\left(\delta v\right)$. The Ollivier--Ricci curvature is defined as:
$$
\kappa_r(\gamma)=\frac{\ell_1(\mu_x,\mu_y)}{\tau(x,y)}-1.
$$
When $\delta$ and $\varepsilon$ are sufficiently small, we have $\tau(x,y)=\delta$, and the Ollivier--Ricci curvature recovers the Ricci curvature up to a factor $\varepsilon^2$. 

\section*{Acknowledgement}
I would like to thank Pavel Kurasov for explaining the history of trace formulas for quantum graphs, and Samuël Borza for helpful discussions on Lorentzian optimal transport and for his valuable comments and corrections.

\bibliographystyle{plain}
\bibliography{Ref}

@article{cry,
  title={Stable polynomials and crystalline measures},
  author={Kurasov, Pavel and Sarnak, Peter},
  journal={Journal of Mathematical Physics},
  volume={61},
  number={8},
  year={2020},
  publisher={AIP Publishing}
}

@article{Zhao2025,
  author        = {Tianhong Zhao},
  title         = {A finite dimensional trace formula},
  year          = {2025},
  eprint        = {2505.14947},
  archivePrefix = {arXiv},
  primaryClass  = {math-ph}
}

@book{Kurasov2024,
  author    = {Pavel Kurasov},
  title     = {Spectral Geometry of Graphs},
  series    = {Operator Theory: Advances and Applications},
  volume    = {293},
  publisher = {Birkh{\"a}user},
  year      = {2024},
  doi       = {10.1007/978-3-662-67872-5}
}

@article{KurasovNowaczyk2005,
  author  = {Kurasov, Pavel and Nowaczyk, Marlena},
  title   = {Inverse spectral problem for quantum graphs},
  journal = {Journal of Physics A: Mathematical and General},
  volume  = {38},
  number  = {22},
  pages   = {4901--4915},
  year    = {2005},
  doi     = {10.1088/0305-4470/38/22/014}
}

@incollection{Roth1984,
  author    = {Roth, Jean-Pierre},
  title     = {Le spectre du Laplacien sur un graphe},
  booktitle = {Th{\'e}orie du Potentiel},
  editor    = {Mokobodzki, Gabriel and Pinchon, Daniel},
  series    = {Lecture Notes in Mathematics},
  volume    = {1096},
  pages     = {521--539},
  publisher = {Springer},
  address   = {Berlin},
  year      = {1984},
  doi       = {10.1007/BFb0100128}
}

@article{BolteEndres2009,
  author  = {Bolte, Jens and Endres, Sebastian},
  title   = {The trace formula for quantum graphs with general self adjoint boundary conditions},
  journal = {Annales Henri Poincar{\'e}},
  volume  = {10},
  pages   = {189--223},
  year    = {2009},
  doi     = {10.1007/s00023-009-0399-7}
}

@article{KottosSmilansky1999,
  author  = {Kottos, Tsampikos and Smilansky, Uzy},
  title   = {Periodic Orbit Theory and Spectral Statistics for Quantum Graphs},
  journal = {Annals of Physics},
  volume  = {274},
  number  = {1},
  pages   = {76--124},
  year    = {1999},
  doi     = {10.1006/aphy.1999.5904}
}

@misc{barton2026ollivierriccicurvaturecausalsets,
  title         = {Ollivier-Ricci Curvature for Causal Sets},
  author        = {Joe Barton and Samu{\"e}l Borza and Jona R{\"o}hrig},
  year          = {2026},
  eprint        = {2606.04910},
  archivePrefix = {arXiv},
  primaryClass  = {math.DG},
  doi           = {10.48550/arXiv.2606.04910}
}

@article{MondinoSuhr2023,
  author  = {Mondino, Andrea and Suhr, Stefan},
  title   = {An optimal transport formulation of the {E}instein equations of general relativity},
  journal = {Journal of the European Mathematical Society},
  volume  = {25},
  number  = {3},
  pages   = {933--994},
  year    = {2023},
  doi     = {10.4171/JEMS/1188},
  url     = {https://doi.org/10.4171/JEMS/1188}
}

\end{document}